\documentclass[10pt,journal,compsoc]{IEEEtran}

\ifCLASSOPTIONcompsoc
\usepackage[nocompress]{cite}
\else
\usepackage{cite}
\fi

\ifCLASSINFOpdf
\usepackage[pdftex]{graphicx}
\else
\fi

\usepackage{amsmath}

\usepackage{url}

\usepackage{amsfonts} 
\usepackage{pgfplots}
\usepackage{subcaption} 

\usepackage{multirow}

\usepackage[linesnumbered,ruled]{algorithm2e}

\begin{document}
	
	\title{A Response to: A Note on “Privacy Preserving $n$-Party Scalar Product Protocol”}
		
	\author{Florian~van Daalen,
		Lianne~Ippel,
		Andre~Dekker,
		and~ Inigo~Bermejo
		\IEEEcompsocitemizethanks{\IEEEcompsocthanksitem  F. van Daalen, I. Bermejo, and A. Dekker are with 	Radiation Oncology (MAASTRO) GROW School for Oncology and Reproduction, University Maastricht Medical Centre+, Maastricht, Netherlands\protect\\
		\IEEEcompsocthanksitem  I. Bermejo is also with 	Data Science Institute, Hasselt University, Hasselt, Belgium\protect\\
		\IEEEcompsocthanksitem L. Ippel is with Statistics Netherlands Heerlen the Netherlands.\protect\\
		\IEEEcompsocthanksitem The views expressed in this paper are those of the authors and do not necessarily reflect the policy of Statistics Netherlands.}
	}
	
	
	\IEEEtitleabstractindextext{%
		\begin{abstract}
			We reply to the comments on our proposed privacy preserving $n$-party scalar product protocol made by Liu. In their comment Liu raised concerns regarding the security and scalability of the $n$-party scalar product protocol. In this reply, we show that their concerns are unfounded and that the $n$-party scalar product protocol is safe for its intended purposes. Their concerns regarding the security are based on a misunderstanding of the protocol. Additionally, while the scalability of the protocol puts limitations on its use, the protocol still has numerous practical applications when applied in the correct scenarios. Specifically within vertically partitioned scenarios, which often involve few parties, the protocol remains practical. In this reply we clarify Liu's misunderstanding. Additionally, we explain why the protocols scaling is not a practical problem in its intended application.
		\end{abstract}
		
		}

	\maketitle

	\IEEEdisplaynontitleabstractindextext

	%

	\section{Introduction}
	\label{sec:introduction}
	In 2023, we proposed a privacy preserving $n$-party scalar product protocol\cite{vanDaalen2023privacy}. The $n$-party scalar product protocol allows the user to calculate a scalar product of vectors spread out over multiple parties in a privacy preserving manner without revealing the content of the vectors. This protocol is an extension of an older protocol proposed by Du and Zhan\cite{du2002building} which was designed for two parties. Recently, Liu has released a preprint\cite{liu2024note} in which it is argued our proposed protocol is insecure and impractical.
	
	First, they claim that our protocol is insecure against a semi-honest server attack. Secondly, they point towards practical concerns regarding the exponential complexity. In this short reply, we will address their concerns and show that the protocol remains a safe and appropriate solution in certain useful scenarios within federated learning.
	
	\begin{algorithm}
		\SetKwInOut{Input}{Input}
		\SetKwInOut{Output}{Output}	
		\SetKw{KwBy}{by}
		
		\underline{nPartyScalarProduct($\mathcal{D}$)} \\
		\Input{ The set $\mathcal{D}$ of diagonal matrices $\bold{D_{1}} .. \bold{D_{n}}$ containing the original vectors owned by the $n$ parties}
		\Output{$\varphi(\bold{D_{1}} \cdot \bold{D_{2}} \cdot..\cdot \bold{D_{n}})$}
		\eIf{$|\mathcal{D}| = 2$}
		{
			return $2$-party scalar product protocol($\mathcal{D}$);
		}
		{		
			\For{$i\gets0$ \KwTo $|\mathcal{D}|$ \KwBy $1$}{
				$\bold{R_{i}} \gets generateRandomDiagonalMatrix()$ \\
			}
			Let $\varphi(\bold{R_{1}} \cdot \bold{R_{2}} \cdot..\cdot \bold{R_{n}}) = r_{1} + r_{2} + … + r_{n}$ \\
			Share $\{\bold{R_{i}},r_{i}\}$ with the $i$'th party for each $i \in [1,n]$ \\
			$v_{2} \gets randomInt()$\\
			$u_1 \gets  \varphi(\prod_{i=2}^{n}\bold{\hat{D}_{i}} \cdot \bold{D_{1}})+ (n-1)\cdot r_{1}- v_{2}$ \\
			\For{$i\gets2$ \KwTo $|\mathcal{D}|$ \KwBy $1$}{
				$u_{i} = u_{i-1} - \newline \varphi((\prod_{x=1}^n {\bf \hat{D}_x}|x\neq i)\cdot {\bf R_i}) \newline + (n-1) \cdot r_{i}$
			}
			$y \gets u_{n}$ \\
			\For{sub-protocol $\in$ determinesub-protocols($\mathcal{D}, \mathcal{R}$)}{
				$y \gets y -$ nPartyScalarProduct($sub-protocol$)
			}
			return $y + v_{2}$		
		}
		\underline{determinesub-protocols($\mathcal{D}, \mathcal{R}$)} \\
		\Input{The set $\mathcal{D}$ of diagonal matrices $\bold{D_{1}} .. \bold{D_{n}}$ of the original protocol. The set $\mathcal{R}$ of random diagonal matrices used in the original protocol}
		\Output{The sets $\mathcal{D}_{sub-protocol}$ for each sub-protocol}
		
		\For{$k\gets2$ \KwTo $|\mathcal{D}| -1$ \KwBy $1$}{
			$uniqueCombinations \gets selectKSizedCombosFromSet(k, \mathcal{D})$ \\
			\For{$selected \in uniqueCombinations$}{
				$sub-protocol \gets  \bold{D_{i}} | i \in selected  + \bold{R_{j}} | j \not \in selected$
				$\mathcal{D}_{sub-protocols} \gets  \mathcal{D}_{sub-protocols} +  sub-protocol$ 	
			}
		}
		
		return 	$\mathcal{D}_{sub-protocols}$
		\caption{The n-party scalar product protocol}\label{pseudocode}
	\end{algorithm}
	
	\section{The privacy preserving $n$-party scalar product protocol}
	Our extension made it possible to use the protocol in an $n$-party scenario. We made this possible by transforming the original vectors into diagonal matrices and using the trace map $\varphi$ of the product of these matrices. Like the original protocol proposed by Du and Zhan, it relies on a trusted third party to generate secret shares. Additionally, our protocol requires sub-protocols to be solved. These sub-protocols require their own secret shares to be generated. However, this cannot be done by the same party that generated the secrets used in the parent protocol. Importantly, Liu's concern relies on the assumption that the secrets are always generated by the same party. Fortunately, since each sub-protocol is at least one party smaller than the parent protocol, it is possible for the party that is not involved in this sub-protocol to take on the role of the trusted third party for this sub-protocol. The pseudo-code of the protocol can be found in algorithm \ref{pseudocode}.
	
	\section{Is the $n$-party protocol vulnerable to a semi-honest server attack?}
	Liu argues that the $n$-party scalar product protocol is vulnerable against a semi-honest server attack. They illustrate their argument using a $3$-party protocol between Alice, Bob, and Claire using sensitive data and a trusted third party called Merlin. This $3$ party protocol will contain a $2$-party sub-protocol of the following form: $\varphi(A*M_{a})$ where $A$ represents the data belonging to Alice, and $M_{a}$ is only known to Merlin and represents the multiplication of the secret shares belong to Bob and Clair. Liu states that in order to solve this sub-protocol Alice will send $\hat{A}=A+R_{a}$ to Merlin. Since Merlin has access to $R_{a}$ this would make it trivial for Merlin to determine $A$.
	
	However, this is incorrect. Alice will not send $\hat{A}=A+R_{a}$ to Merlin. Instead Bob (or Claire) will generate a new secret share $R_{a}^{sub-protocol}$ and Alice will share $\hat{A^{sub-protocol}} = A + R_{a}^{sub-protocol}$ with Merlin. Since Merlin does not know $R_{a}^{sub-protocol}$ it cannot determine $A$. This means the protocol is secure and can be continued to be utilized without any problems. This addresses Liu's concern about the insecurity of the protocol.
	
	\section{Time complexity}
	According to Liu, the $n$-party scalar product protocol has an exponential complexity in the number of sub-protocols. Liu argues that because of this, the protocol cannot be used in practice. It is indeed correct that the protocol has an exponential complexity and this is acknowledged as a limitation in the original paper. As such the $n$-party scalar product protocol is best used in scenarios with relatively few parties, for example a research project using federated learning that involves multiple hospitals in a horizontally partitioned scenario, or when a hospital wishes to work together with an insurance agency in vertically partitioned scenario. Our proposed method is especially relevant in a vertical partitioned scenario, as such scenarios rarely have a large number of parties. As explained in the original publication, one of the main intended uses of the protocol is to determine the size of a subset of the population which fulfill certain criteria, even when the relevant attributes are spread across different parties. As such, the protocol is largely intended for vertically partitioned scenarios. Additionally, since publishing the $n$-party scalar product protocol we have published multiple follow up papers illustrating that the protocol can be used in such situations and has an acceptable runtime to be used in practice\cite{vanDaalen2023federated, vanDaalen2024vertibayes}. 
	
	\section{Conclusion}
	In this short reply, we have shown that Liu's concerns with the $n$-party scalar product protocol are  unfounded and that the protocol is safe. We believe that the security concerns arose from a misunderstanding of the protocol. Additionally, we have shown that, while the protocol does indeed scale poorly in the number of parties, this is not a problem in practice as the protocol is intended for vertically split scenarios in which the number of parties naturally remains low. We would like to thank Liu for their efforts in helping us improve the protocol and clear up any potential misunderstandings. We hope this addresses any concerns that may have arisen considering the protocol.

	\bibliographystyle{IEEETran}
	\bibliography{reply}
\end{document}